\documentclass[amsmath,aps,showpacs,a4paper,10pt]{revtex4}

 \usepackage{epsf}
 \usepackage{graphicx}    

 \usepackage{amssymb}     

\begin{document}

 \newcommand{\be}[1]{\begin{equation}\label{#1}}
 \newcommand{\ee}{\end{equation}}
 \newcommand{\bea}{\begin{eqnarray}}
 \newcommand{\eea}{\end{eqnarray}}
 \def\disp{\displaystyle}

 \begin{titlepage}

 \begin{flushright}
 arXiv:2103.12696
 \end{flushright}

 \title{\Large \bf Inverse Chameleon Mechanism and Mass
 Limits\\ for Compact Stars}

 \author{Hao~Wei\,}
 \email[\,Corresponding author;\ email address:\ ]{haowei@bit.edu.cn}
 \affiliation{School of Physics,
 Beijing Institute of Technology, Beijing 100081, China}

 \author{Zhong-Xi~Yu\,}
 \email[\,email address:\ ]{547410406@qq.com}
 \affiliation{School of Physics,
 Beijing Institute of Technology, Beijing 100081, China}

 \begin{abstract}\vspace{1cm}
 \centerline{\bf ABSTRACT}\vspace{2mm}
 As is well known, there are various mass limits for compact stars. For
 example, the maximum mass for non-rotating white dwarfs is given by the
 famous Chandrasekhar limit about $1.4 M_\odot$ (solar masses). Although
 the mass limit for neutron stars is not so clear to date, one of the
 widely accepted values is about $2.1 M_\odot\,$. Recently, challenges
 to these mass limits appeared. Motivated by the super-Chandrasekhar mass
 white dwarfs with masses up to $2.4 \sim 2.8 M_\odot\,$, and compact
 objects (probably neutron stars) in the mass gap (from $2.5 M_\odot$ or
 $3 M_\odot$ to $5 M_\odot$) inferred from gravitational waves detected by
 LIGO/Virgo in the third observing run (O3), we reconsider the mass limits
 for compact stars in the present work. Without invoking strong magnetic
 field and/or exotic equation of state (EOS), we try to increase the mass
 limits for compact stars in modified gravity theory. In this work, we
 propose an inverse chameleon mechanism, and show that the fifth-force
 mediated by the scalar field can evade the severe tests on earth, in solar
 system and universe, but manifest itself in compact stars such as white
 dwarfs and neutron stars. The mass limits for compact stars in the inverse
 chameleon mechanism can be easily increased to $3 M_\odot\,$, $5 M_\odot$
 or even larger. We argue that the inverse chameleon mechanism might be
 constrained by the observations of exoplanets orbiting compact stars (such
 as white dwarfs and neutron stars), and gravitational waves from the
 last stage of binary compact star coalescence.
 \end{abstract}


 \maketitle

 \end{titlepage}

 \renewcommand{\baselinestretch}{1.0}


\section{Introduction}\label{sec1}

As is well known, when a massive star ends, the stellar core remnant
 might form a compact object such as white dwarf, neutron star, and
 black hole~\cite{Camenzind:2007,Shapiro:2004,Glendenning:1996}. At
 this stage, fusion reactions in the star stopped, and hence gravitational
 collapse must take place. If the mass of star is less than about $10
 M_\odot$ (solar masses), a white dwarf will be formed when gravitational
 collapse is eventually balanced by electron degeneracy pressure. As is
 well known, there is a maximum mass for non-rotating white dwarfs, namely
 the Chandrasekhar limit~\cite{Chandrasekhar:1931ih,Chandrasekhar:1931ftj,
 Chandrasekhar:1935zz,Chandrasekhar:1939}, whose currently accepted value
 is about $1.4 M_\odot\,$. A white dwarf with a mass larger than the
 Chandrasekhar limit is subject to further gravitational collapse, and
 hence evolves into a neutron star. Neutron stars are supported against
 further collapse by neutron degeneracy pressure and repulsive nuclear
 forces. There is also a maximum mass for non-rotating neutron
 stars~\cite{Camenzind:2007,Shapiro:2004,Glendenning:1996}, namely the
 Tolman-Oppenheimer-Volkoff (TOV) limit. However, this mass limit is not
 so clear to date~\cite{Ozel:2016oaf,Ozel:2012ax,Chamel:2013efa}. One of
 the widely accepted values is about $2.1 M_\odot$~\cite{Ozel:2016oaf,
 Ozel:2012ax,Chamel:2013efa}. A recent estimate puts the upper limit at
 $2.16 M_\odot$~\cite{Rezzolla:2017aly}. To date, the maximum observed
 mass of neutron star is about $2.14 M_\odot$ for PSR J0740+6620 discovered
 in September 2019~\cite{Cromartie:2019kug}. If exotic equation of state
 (EOS) is allowed, this mass limit might be $2.2 M_\odot\,$, $2.5 M_\odot$
 or even higher (see e.g.~\cite{Burrows:2017,Kalogera:1996ci}).
 A neutron star with a mass larger than the TOV limit is subject to further
 gravitational collapse, and then forms a black hole. However, hypothetical
 intermediate-mass stars such as quark star, boson star, electroweak star
 and gravastar might exist between neutron star and black hole, although
 none of them has been discovered to date. Of course, there must also be
 mass limits for these hypothetical compact stars, beyond which they
 collapse into a black hole.

Recently, challenges to these mass limits appeared. It is widely accepted
 that type Ia supernovae (SNIa) are explosions of carbon-oxygen white
 dwarfs. For the first time, the progenitor of a very bright SNIa, namely
 SNLS-03D3bb (SN 2003fg), was found to be a super-Chandrasekhar mass white
 dwarf in 2006~\cite{Howell:2006vn}. In fact, the mass of this white dwarf
 is about $2.1 M_\odot\,$, highly exceeding the Chandrasekhar mass limit
 (about $1.4 M_\odot$). Later, more super-Chandrasekhar mass white dwarfs
 were found to be the progenitors of very bright SNIa, for example, SN
 2006gz, SN 2007if, and SN 2009dc, as summarized in e.g.~\cite{Das:2013gd,
 Hachisu:2011jv}. These super-Chandrasekhar mass white dwarfs have masses
 up to $2.4 \sim 2.8 M_\odot\,$. In addition, a super-Chandrasekhar nucleus
 of the planetary nebula Henize 2-428 with a combined mass of $1.76 M_\odot$
 was found~\cite{Santander-Garcia:2015}. To date, fewer than ten candidates
 of super-Chandrasekhar mass white dwarfs are under consideration, for
 example, the progenitors of SN 2012dn, SN 2011aa, SN 2011hr, SN 2004gu,
 LSQ12gdj, and iPTF13asv, as summarized in e.g.~\cite{Hsiao:2020whc,
 Brown:2014jua}. These super-Chandrasekhar mass white dwarfs clearly require
 a new theoretical mass limit for white dwarfs.

On the other hand, direct detections of gravitational waves (GWs) become
 available since September 2015
 \cite{Abbott:2016blz,TheLIGOScientific:2016src}. In the LIGO/Virgo
 classification, a ``\,MassGap\,'' system refers to a binary system with at
 least one compact object whose mass is in the range of $3\sim 5 M_\odot$
 \cite{MassGap}. A natural question is that if a MassGap GW event has been
 detected, what is the compact object with a mass between $3 M_\odot$ and
 $5 M_\odot$? It might be a large mass neutron star, or a small mass black
 hole. The difference is whether there is electromagnetic counterpart or
 not. On 16 December 2019, LIGO/Virgo detected a GW event S191216ap, and
 initially classified it as MassGap with a $>99\%$ probability (GCN
 circular 26454)~\cite{S191216ap:GCN,S191216ap:gracedb}. Soon, the IceCube
 Collaboration claimed that a neutrino counterpart associated with S191216ap
 was found (GCN circular 26460) \cite{S191216ap:GCN}. Then, the HAWC
 Collaboration claimed that a gamma-ray counterpart associated with
 S191216ap was also found (GCN circular 26472) \cite{S191216ap:GCN}. Since
 two electromagnetic counterparts were claimed, the compact object in
 MassGap cannot be a black hole, and it is probably a neutron star with a
 mass larger than $3 M_\odot\,$, exceeding the theoretical mass limits for
 neutron stars mentioned above. Unfortunately, LIGO/Virgo changed the
 classification of S191216ap to BBH ($>99\%$) on 19 December 2019 (GCN
 circular 26570)~\cite{S191216ap:GCN,S191216ap:gracedb}, and then the story
 ended. Anyway, this motivates us to reconsider the theoretical mass limit
 for neutron stars. We should be ready in advance for a MassGap neutron star
 ($M>3M_\odot$) associated with electromagnetic counterparts in the future.

Another very important GW event is GW190814
 \cite{S190814bv:GCN,S190814bv:gracedb,Abbott:2020khf}. LIGO/Virgo initially
 classified this event as MassGap ($>99\%$) on 14 August 2019 (GCN circular
 25324) \cite{S190814bv:GCN,S190814bv:gracedb}, and then changed it to NSBH
 ($>99\%$) on the next day (GCN circular 25333) \cite{S190814bv:GCN,
 S190814bv:gracedb}. After detailed analyses, LIGO/Virgo found
 in~\cite{Abbott:2020khf} that it came from the coalescence of
 a $23 M_\odot$ black hole with a $2.6 M_\odot$ compact object (note that
 in the press release~\cite{GW190814:pr}, LIGO/Virgo still classified the
 $2.6 M_\odot$ compact object in the mass gap). Because no black hole with
 a mass less than $5 M_\odot$ was observed before, if the $2.6 M_\odot$
 compact object is a black hole, could it be a primordial black hole? As
 mentioned above, the maximum observed mass of neutron star is about
 $2.14 M_\odot$ to date \cite{Cromartie:2019kug}. The mass of this compact
 object ($2.6 M_\odot$) also well exceeds the widely accepted mass limit
 for neutron stars mentioned above (about $2.16 M_\odot$). Unfortunately,
 no electromagnetic counterpart associated with GW190814 was claimed
 \cite{S190814bv:GCN} (but see also e.g.~\cite{Wei:2019wxd}). So, none can
 tell whether this $2.6 M_\odot$ compact object is a neutron star or not.
 Anyway, this motivates us again to reconsider the theoretical mass limit
 for neutron stars.

In the literature, there are various scenarios to increase the mass limits
 for compact stars such as white dwarfs and neutron stars. Strong magnetic
 field and/or exotic EOS are frequently invoked in many scenarios (see
 e.g.~\cite{Das:2014ssa,Das:2012ai,Roy:2019nja,Zou:2015vxa,Shah:2020}
 and~\cite{Kalogera:1996ci,Bombaci:1996,Zdunik:2012dj,Godzieba:2020tjn,
 Studzinska:2016ofb,Chamel:2012ea}). Another type of scenarios is to
 consider compact stars in modified gravity theories. If gravitational
 force is weaken (with respect to general relativity), the mass limits
 for compact stars can accordingly increase, as expected. We refer to
 e.g.~\cite{Olmo:2019flu} for a comprehensive review. However, it is worth
 noting that in modified gravity theories, gravitational force is modified
 on all scales, not only in compact stars but also in solar system and
 universe. In fact, many modified gravity theories, in which the mass limits
 for compact stars could be considerably increased, will significantly
 deviate from general relativity (GR), and hence they are difficult to
 simultaneously evade the severe tests on earth and in solar system (as
 well as the cosmological tests).

The key is to make gravity environment-dependent. One of this
 kind of modified gravity theories is the well-known chameleon mechanism
 \cite{Khoury:2003aq,Khoury:2003rn,Brax:2004qh,Gubser:2004uf,Khoury:2013yya,
 Wei:2004rw}, which was proposed mainly for cosmology. This mechanism can
 hide dark energy (played by a scalar field coupling to matter, namely the
 so-called chameleon field) on earth and in solar system, but show it on
 cosmological and galactic scales. The mass of scalar field depends on the
 ambient matter density. On earth and in solar system, where the matter
 density is high, the scalar field is massive, and hence the fifth-force
 range is short enough to evade the severe tests on earth and in solar
 system. On cosmological and galactic scales, where the matter density is
 low, the mass of scalar field is light, and hence the fifth-force
 range is long enough to drive the cosmic acceleration or the evolution of
 the fine-structure ``\,constant\,''. Unfortunately, this chameleon
 mechanism cannot be used to increase the mass limits for compact stars.
 The matter density is very high in compact stars, and hence the range of
 fifth-force mediated by the scalar field is too short to manifest itself.

In the present work, we try to invert the chameleon mechanism. In our
 inverse chameleon mechanism, we will show the fifth-force mediated by the
 scalar field in compact stars, and hide it on earth, in solar system and
 universe. So, the mass limits for compact stars can be significantly
 increased in the inverse chameleon mechanism, and simultaneously evade
 the severe tests on earth and in solar system (as well as the cosmological
 tests).

The rest of this paper is organized as follows. In Sec.~\ref{sec2}, we
 briefly review the key points of chameleon mechanism. In Secs.~\ref{sec3}
 and \ref{sec4}, we propose our inverse chameleon mechanism. We present
 the solutions for a compact object, and show that the fifth-force mediated
 by the scalar field can evade the severe tests on earth, in solar system
 and universe, but manifest itself in compact stars such as white dwarfs and
 neutron stars. In Sec.~\ref{sec5}, we derive the new mass limits for white
 dwarfs and other compact stars in the inverse chameleon mechanism.
 In fact, they can be easily increased, exceeding the mass gap, namely
 $M > 3 M_\odot$ or even larger. In Sec.~\ref{sec6}, some brief concluding
 remarks are given.


\section{The key points of chameleon mechanism}\label{sec2}

At first, we briefly review the key points of chameleon
 mechanism, following e.g.~\cite{Khoury:2003aq,Khoury:2003rn,Brax:2004qh,
 Gubser:2004uf,Khoury:2013yya,Wei:2004rw}. In the Einstein frame, the
 canonical scalar field $\phi$ (namely the chameleon field) is
 governed by the action
 \be{eq1}
 S=\int d^4 x\,\sqrt{-g}\left[\frac{M_{pl}^2}{2}\,R-\frac{1}{2}\left(
 \partial\phi\right)^2-V(\phi)\right]+S_m(\,g^{\rm\, J})\,,
 \ee
 in which matter fields described by $S_m$ couple to $\phi$ through
 the conformal factor $A(\phi)$ implicit in the Jordan-frame
 metric~\cite{Khoury:2013yya}
 \be{eq2}
 g^{\rm\, J}_{\mu\nu}=A^2(\phi)\; g_{\mu\nu}\,,
 \ee
 and $M_{pl}\equiv (8\pi G)^{-1/2}$ is the reduced Planck mass, $g$ is
 the determinant of the metric $g_{\mu\nu}$, $R$ is the Ricci scalar. We
 use the units $\hbar=c=1$, and the metric convention $(-,\,+,\,+,\,+)$.
 In principle, one can allow different couplings to the various matter
 fields through $g^{{\rm\, J}\,(i)}_{\mu\nu}=A_i^2(\phi)\; g_{\mu\nu}$,
 explicitly violating the equivalence principle, as in e.g.
 \cite{Khoury:2003aq,Khoury:2003rn,Brax:2004qh,Wei:2004rw}. For
 simplicity, following e.g.~\cite{Khoury:2013yya}, we only consider the
 simplest case of a universal coupling in this work, without violating
 the equivalence principle. From the action~(\ref{eq1}), the equation
 of motion for $\phi$ is given by~\cite{Khoury:2013yya,Wang:2012kj}
 \be{eq3}
 \Box\phi=V_{,\phi}+A_{,\phi}\,\rho=V_{{\rm eff},\phi}\,,
 \ee
 where $\Box$ is the d'Alembertian, $f_{,\phi}$ denotes the
 derivative of any function $f$ with respect to $\phi$, and the
 effective potential is defined by
 \be{eq4}
 V_{\rm eff}(\phi)=V(\phi)+A(\phi)\,\rho\,.
 \ee
 The matter density $\rho$ is related to the Einstein-frame matter density
 $\rho_{\rm E}$ and the Jordan-frame matter density $\rho_{\rm J}$ by
 $\rho=\rho_{\rm E}/A=A^3\rho_{\rm J}$~\cite{Khoury:2013yya}, so that $\rho$
 is conserved in the Einstein frame~\cite{Khoury:2003aq,Khoury:2003rn,
 Brax:2004qh,Gubser:2004uf,Khoury:2013yya,Wei:2004rw}. On the other
 hand, the acceleration of a test particle is influenced by the
 scalar field according to~\cite{Khoury:2013yya,Wang:2012kj}
 \be{eq5}
 \boldsymbol{a}=-\nabla \Phi_{\rm N}-\frac{d\ln A(\phi)}{d\phi}\,\nabla\phi
 =-\nabla\left(\Phi_{\rm N}+\ln A(\phi)\right)\,,
 \ee
 where $\Phi_{\rm N}$ is the (Einstein-frame) Newtonian potential, which
 satisfies~\cite{Khoury:2013yya,Wang:2012kj}
 \be{eq6}
 \nabla^2 \Phi_{\rm N}=4\pi G\rho_{\rm E}=4\pi G A\rho\,.
 \ee

In the chameleon mechanism, an exponential
 coupling is usually considered, i.e.~\cite{Khoury:2003aq,
 Khoury:2003rn,Brax:2004qh,Gubser:2004uf,Khoury:2013yya,Wei:2004rw}
 \be{eq7}
 A(\phi)=\exp\left(\beta\phi/M_{pl}\right)\,,
 \ee
 where $\beta\sim {\cal O}(1)$ is a dimensionless constant. Clearly,
 $A(\phi)$ is monotonically increasing. The potential $V(\phi)$ is assumed
 to be of the runaway form, so that it is monotonically decreasing. The
 fiducial example is an inverse power-law potential~\cite{Khoury:2003aq,
 Khoury:2003rn,Khoury:2013yya}
 \be{eq8}
 V(\phi)={\cal M}^4\,({\cal M}/\phi)^n\,,
 \ee
 where $\cal M$ has units of mass, and $n$ is a positive constant. So,
 the effective potential $V_{\rm eff}$ can develop a minimum at some
 finite field values $\phi_{\rm min}$ in the presence of background matter
 density. It is easy to find $\phi_{\rm min}\propto\rho^{-1/(n+1)}$
 \cite{Khoury:2013yya} by requiring $V_{\rm eff,\phi}(\phi_{\rm min})=0$.
 Note that we only consider the case of $\phi\ll M_{pl}$ throughout this
 work, following e.g.~\cite{Khoury:2003aq,Khoury:2003rn,Brax:2004qh,
 Gubser:2004uf,Khoury:2013yya,Wei:2004rw}. The mass of small fluctuations
 about the minimum at $\phi_{\rm min}$ for the canonical scalar field
 $\phi$ is defined as usual~\cite{Khoury:2003aq,Khoury:2003rn,Brax:2004qh,
 Gubser:2004uf,Khoury:2013yya,Wei:2004rw}
 \be{eq9}
 m_\phi^2\equiv V_{\rm eff,\phi\phi}(\phi_{\rm min})\geq 0\,.
 \ee
 Noting $\rho$ in the effective potential given by Eq.~(\ref{eq4}), the
 originally massless scalar field $\phi$ acquires a mass depending on the
 local matter density. Substituting $\phi_{\rm min}$ into Eq.~(\ref{eq9}),
 and noting $\phi\ll M_{pl}$, we find that
 $m_\phi^2\propto\rho^{(n+2)/(n+1)}$~\cite{Khoury:2013yya} is an increasing
 function of the background density. So, on earth and in solar system, where
 the matter density is high, the scalar field is massive, and hence the
 fifth-force range is short enough to evade the severe tests on earth and in
 solar system. On cosmological and galactic scales, where the matter density
 is low, the mass of scalar field is light, and hence the fifth-force range
 is long enough to drive the cosmic acceleration or the evolution of the
 fine-structure ``\,constant\,''. We refer to e.g.~\cite{Khoury:2003aq,
 Khoury:2003rn,Gubser:2004uf} for the detailed magnitude analyses to evade
 the tests on earth and in solar system. The key point is to require the
 range of fifth-force mediated by the scalar field $\phi$ in the atmosphere
 $m_{\rm atm}^{-1}\lesssim {\cal O}(1\,{\rm mm})$~\cite{Khoury:2003aq,
 Khoury:2003rn,Gubser:2004uf}. For $n$ and $\beta$ of order unity, it can be
 translated into a constraint on the scale $\cal M$, namely ${\cal M}
 \lesssim 10^{-3}\,{\rm eV}\sim (1\,{\rm mm})^{-1}$~\cite{Khoury:2003aq,
 Khoury:2003rn}.

The screening of fifth-force mediated by the scalar field $\phi$ can also
 be seen from the solutions for a compact object~\cite{Khoury:2003aq,
 Khoury:2003rn}. The key is the so-called ``\,thin-shell\,'' effect. We
 refer to \cite{Khoury:2003aq,Khoury:2003rn} for the explicit solutions.
 The exterior solution for a compact object having the thin-shell effect
 is suppressed by a factor $3\Delta R_c/R_c\ll 1$ with respect to the
 exterior solution for a compact object without the thin-shell effect.
 This effect can be understood from the physical picture following e.g.
 \cite{Khoury:2003rn,Khoury:2013yya}. If the object is sufficiently
 massive such that deep inside the object the scalar field minimizes
 the effective potential for the interior density, the mass of scalar
 field is relatively large inside the object, and hence the fifth-force
 range is relatively short. Thus, the contribution from the core to the
 exterior profile is significantly suppressed. Only the contribution
 from a thin shell beneath the surface contributes considerably to the
 exterior profile~\cite{Khoury:2013yya}. This is the physical reason
 of the thin shell effect.

In the literature, there are many interesting works used the chameleon
 mechanism. Of course, most of them concern cosmology. Unfortunately, the
 chameleon mechanism cannot be used to increase the mass limits for compact
 stars, as mentioned in Sec.~\ref{sec1}. The matter density is very high in
 compact stars, and hence the range of fifth-force mediated by the scalar
 field is too short to manifest itself. Therefore, we should find a way out.


\section{Inverse chameleon mechanism}\label{sec3}


\subsection{The ingredients of inverse chameleon mechanism}\label{sec3a}

We try to invert the chameleon mechanism. Since a canonical
 scalar field (akin to quintessence) is used in the chameleon
 mechanism, we instead consider a non-canonical scalar field (akin to
 phantom)~\cite{Caldwell:1999ew} in our inverse chameleon mechanism. As is
 well known, in cosmology, phantom is almost the inverse of quintessence.
 While the kinetic energy term of quintessence is positive, it is negative
 instead in the case of phantom. So, the behaviors of phantom
 and quintessence are almost inverse. For example, in cosmology where
 $\phi$ depends only on the time $t$, phantom rests at the maximum of
 its potential, while quintessence rests at the minimum. Naively, let
 us begin our inverse chameleon mechanism with the action
 \be{eq10}
 S=\int d^4 x\,\sqrt{-g}\left[\frac{M_{pl}^2}{2}\,R+\frac{1}{2}\left(
 \partial\phi\right)^2-V(\phi)\right]+S_m(\,g^{\rm\, J})\,,
 \ee
 where $\phi$ is a non-canonical scalar field (akin to phantom) instead, and
 $g^{\rm\, J}_{\mu\nu}$ takes the same form of Eq.~(\ref{eq2}). The sign of
 $(\partial\phi)^2$ term is opposite to the one in Eq.~(\ref{eq1}). So, the
 equation of motion for $\phi$ is given by
 \be{eq11}
 -\Box\phi=V_{,\phi}+A_{,\phi}\,\rho=V_{{\rm eff},\phi}\,,
 \ee
 which is also opposite to Eq.~(\ref{eq3}), while $V_{\rm eff}$ takes
 the same form of Eq.~(\ref{eq4}). Of course, the acceleration
 of a test particle influenced by the scalar field also takes the same
 form of Eq.~(\ref{eq5}), while Eq.~(\ref{eq6}) still holds for
 the (Einstein-frame) Newtonian potential $\Phi_{\rm N}$.

Our goal is to make the mass of the scalar field small (large) when the
 ambient matter density $\rho$ is large (small), in opposite to the
 chameleon mechanism. Unlike a canonical scalar field (akin to
 quintessence), the mass of a non-canonical scalar field (akin to phantom)
 is defined about the maximum at $\phi_{\rm max}\,$, i.e.
 \be{eq12}
 m_\phi^2\equiv -V_{\rm eff,\phi\phi}(\phi_{\rm max})\geq 0\,,
 \ee
 since $V_{\rm eff,\phi\phi}$ is negative at the maximum of the effective
 potential. Naively, we consider
 \be{eq13}
 V(\phi)=V_0\,\phi^s\,,
 \quad\quad A(\phi)=\exp\left(\beta\phi/M_{pl}\right)\,,
 \ee
 where $V_0$ has units of $[\,{\rm mass}\,]^{4-s}$, and $\beta\sim {\cal O}
 (1)$ is a dimensionless constant. Note that we only consider the case of
 $\phi\ll M_{pl}$ throughout this work, following e.g.~\cite{Khoury:2003aq,
 Khoury:2003rn,Brax:2004qh,Gubser:2004uf,Khoury:2013yya,Wei:2004rw}.
 Requiring $V_{{\rm eff},\phi}(\phi_{\rm max})=0$, we find
 \be{eq14}
 \phi_{\rm max}=\left(-\frac{\beta\rho}{sV_0 M_{pl}}\right)^{1/(s-1)}\,.
 \ee
 To be a maximum, it is required that
 \be{eq15}
 V_{{\rm eff},\phi\phi}(\phi_{\rm max})=V_0\,s(s-1)\left(
 -\frac{\beta\rho}{sV_0 M_{pl}}\right)^{(s-2)/(s-1)}\leq 0\,.
 \ee
 To avoid complex number, $-\beta\rho/(sV_0 M_{pl})>0$ is required. So,
 we have $V_0\,s(s-1)\leq 0$. Noting that
 \be{eq16}
 m_\phi^2=-V_{{\rm eff},\phi\phi}(\phi_{\rm max})
 \propto\rho^{\,(s-2)/(s-1)}\,,
 \ee
 $(s-2)/(s-1)<0$ is required to make $m_\phi$ being a decreasing function
 of $\rho$. So, we should set
 \be{eq17}
 1<s<2\,, \quad\quad V_0\leq 0\,, \quad\quad \beta>0\,,
 \ee
 in the inverse chameleon mechanism. One might worry about the potential
 $V(\phi)=V_0\,\phi^s$ in Eq.~(\ref{eq13}), since it might become complex
 number for $\phi<0$. To extend the relevant range of $\phi$ to the
 negative region, we can instead use $V(\phi)=V_0\left|\phi\right|^s$,
 which is still the same $V(\phi)=V_0\,\phi^s$ for $\phi\geq 0$. But for
 $\phi<0$, noting that $V_0\leq 0$, $V(\phi)=V_0\,(-\phi)^s$ is
 monotonically decreasing as $\phi\to -\infty$. Because $A(\phi)$ is
 also monotonically decreasing as $\phi\to -\infty$, the maximum of
 $V_{\rm eff}$ certainly does not appear in the region of $\phi<0$. In
 fact, noting Eqs.~(\ref{eq17}) and (\ref{eq14}), $\phi_{\rm max}>0$
 always. So, it is safe to use $V(\phi)=V_0\,\phi^s$ in the range of
 $0\leq\phi\ll M_{pl}$ (note that a minimum will develop for
 $\phi\gtrsim M_{pl}$ beyond our scope $\phi\ll M_{pl}$).


\subsection{Constraints on model parameters}\label{sec3b}

In the inverse chameleon mechanism, the mass of the scalar field $m_\phi$
 is a decreasing function of the local matter density $\rho$, as shown in
 Eq.~(\ref{eq16}) with $1<s<2$. So, $m_\phi$ can be very small in compact
 stars such as white dwarfs and neutron stars where the matter density is
 very high, and hence the fifth-force range is long enough (even far beyond
 the radius of compact star) to manifest itself. On the contrary, $m_\phi$
 can be large on earth, in solar system and universe where the matter
 density is relatively low, so that the fifth-force range
 is short enough to evade the severe tests.

Noting $V_0\leq 0$, we recast the potential $V(\phi)$ as
 \be{eq18}
 V(\phi)=V_0\,\phi^s=-{\cal M}^4\,(\phi/{\cal M})^s\,,
 \ee
 where $\cal M$ has units of mass. In this case, we have
 \be{eq19}
 m_\phi^{-1}=\left(s(s-1)\right)^{-1/2} {\cal M}^{\,(s-4)/(2(s-1))}
 \left(\frac{\beta\rho}{sM_{pl}}\right)^{(2-s)/(2(s-1))}\,,
 \ee
 which characterizes the fifth-force range. Note that the mean densities of
 atmosphere, earth and sun are $\rho_{\rm atm}\simeq 1.2\times 10^{-3}\,{\rm
 g/cm^3}$, $\rho_\oplus\simeq 5.514\,{\rm g/cm^3}$ and $\rho_\odot\simeq
 1.408\,{\rm g/cm^3}$, respectively. The densest object on earth is the
 metal Osmium (Os) with $\rho_{\rm Os}=22.59\,{\rm g/cm^3}$. For magnitude
 estimate, it is convenient to simply use $\rho_{\rm metal}\sim {\cal O}(10
 \,{\rm g/cm^3})$. Noting Eq.~(\ref{eq19}) with $1<s<2$, $m_\phi^{-1}$ is a
 increasing function of $\rho$. If the fifth-force range is short enough to
 evade the severe tests for metal, it holds on earth, in solar system and
 universe where $\rho<\rho_{\rm metal}$. Similarly, if the fifth-force
 range is long enough to manifest itself for white dwarfs ($\rho_{\rm WD}
 \sim 10^6 \,{\rm g/cm^3}$), it holds for all compact stars with much
 higher densities.

Let us find the constraints on model parameters. It is worth noting that
 \be{eq20}
 m_{\rm WD}^{-1}\sim m_{\rm metal}^{-1}\left(\rho_{\rm WD}/\rho_{\rm metal}
 \right)^{(2-s)/(2(s-1))}\sim m_{\rm metal}^{-1}\cdot 10^{\,5(2-s)/(2(s-1))}\,,
 \ee
 which increases as $s\to 1$. According to e.g.~\cite{Khoury:2003aq,
 Khoury:2003rn,Brax:2004qh,Gubser:2004uf,Khoury:2013yya,Wei:2004rw},
 $m_\phi^{-1}\lesssim {\cal O}(1\,{\rm mm})$ is enough to evade the
 fifth-force tests on earth and in solar system. So, we
 consider $m_{\rm metal}^{-1}\sim 1\,{\rm mm}$ for magnitude estimate.
 In this case, $m_{\rm WD}^{-1}\sim 10^{33/2}\,{\rm km}$, $10^4\,{\rm km}$,
 $10^{-1/6}\,{\rm km}$, $10^{-9/4}\,{\rm km}$, $10^{-7/2}\,{\rm km}$ for
 $s=1.1$, $1.2$, $1.3$, $1.4$, $1.5$, respectively. Noting that the typical
 radius of white dwarfs $\sim {\cal O}(10^3\,{\rm km})$, it is suitable to
 use $1<s\lesssim 1.2$ to manifest the fifth-force in white dwarfs. On the
 other hand, for earth, sun, diamond, ceramics, silicon, and metals of
 $\rho\sim {\cal O}(1\,{\rm g/cm^3})$, we have
 $m_\oplus^{-1}\sim m_\odot^{-1}\sim m_{\rm light\; metal}^{-1}\sim
 m_{\rm metal}^{-1}\cdot 10^{(s-2)/(2(s-1))}\lesssim 10^{-2}\,m_{\rm
 metal}^{-1}\sim 10^{-2}\,{\rm mm}$ for $1<s\lesssim 1.2$. Similarly,
 $m_{\rm atm}^{-1}\sim m_{\rm metal}^{-1}\cdot 10^{2(s-2)/(s-1)}\lesssim
 10^{-8}\,m_{\rm metal}^{-1}\sim 10^{-8}\,{\rm mm}$ for $1<s\lesssim 1.2$.
 Let us turn to the scale $\cal M$. Noting Eq.~(\ref{eq19}), $m_\phi^{-1}
 \lesssim {\cal O}(1\,{\rm mm})$ to evade the fifth-force tests on earth and
 in solar system can be translated to ${\cal M}\gtrsim 10\,{\rm mm}^{-1}
 \sim 2\times 10^{-3}\,{\rm eV}$ for $s=1.2$, $\beta=1/2$, and $\rho\sim
 20\,{\rm g/cm^3}$. Of course, one can instead take a much conservative
 value, e.g. $m_\phi^{-1}\lesssim {\cal O}(1\,{\rm \mu m})$, to evade the
 fifth-force tests on earth and in solar system. We consider
 $m_{\rm metal}^{-1}\sim 1\,{\rm \mu m}$ for magnitude estimate. In this
 case, $s=1.2$ is not enough to make $m_{\rm WD}^{-1}$ larger than the
 typical radius of white dwarfs. But $s=1.1$ is certainly enough, since the
 corresponding $m_{\rm WD}^{-1}\sim 10^{27/2}\,{\rm km}$. On the other
 hand, we find that ${\cal M}\gtrsim 15\,{\rm mm^{-1}}\sim 3\times 10^{-3}
 \,{\rm eV}$ for $s=1.1$, $\beta=1/2$, and $\rho\sim 10\,{\rm g/cm^3}$.
 In summary, we can always use a $s$ close to 1 and a higher ${\cal M}
 \gtrsim {\cal O}(10\,{\rm mm^{-1}})\sim {\cal O}(10^{-3}\,{\rm eV})$ to
 evade the tests on earth, in solar system and universe, but manifest the
 fifth-force in compact stars such as white dwarfs and neutron stars.


\section{The solutions for a compact object}\label{sec4}


\subsection{The qualitative description}\label{sec4a}

\begin{center}
 \begin{figure}[tb]
 \centering
 \vspace{-8mm}  
 \includegraphics[width=0.8\textwidth]{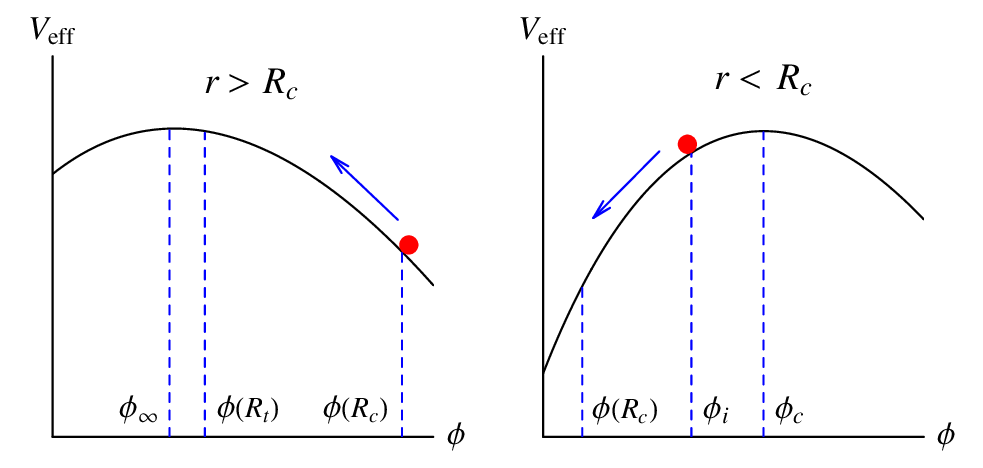}
 \caption{\label{fig1} The effective potential $V_{\rm eff}$ for a compact
 object of radius $R_c$ is discontinuous at $r=R_c\,$, since the matter
 density $\rho=\rho_c$ for $r<R_c$ (right panel), and $\rho=\rho_\infty$
 for $r>R_c$ (left panel). The scalar field $\phi(r)$ (red solid balls)
 rolls from $\phi_i$ to $\phi_\infty$, as $r$ runs from $0$ to $\infty$.
 Since $\phi_\infty<\phi_c$ for $\rho_\infty<\rho_c$ usually, it is more
 convenient to put the $r>R_c$ panel on the left side of the $r<R_c$ panel.
 The plots are not to scale. See Sec.~\ref{sec4} for details.}
 \end{figure}
 \end{center}

Here, we derive the approximate solutions of the scalar field $\phi$
 in the inverse chameleon mechanism for a spherical homogeneous
 isolated compact object, similar to the case of chameleon
 mechanism~\cite{Khoury:2003aq,Khoury:2003rn}. Let the radius, density
 and total mass of this object be $R_c\,$, $\rho_c$ and $M_c=4\pi\rho_c
 R_c^3/3$, respectively. We assume that it is immersed in a background
 of homogeneous density $\rho_\infty$ (less than $\rho_c$ usually), similar
 to objects in the atmosphere, earth in the solar plasma, sun and (compact)
 stars in the interstellar gas, galaxies in the intergalactic medium. So,
 the equation of motion in Eq.~(\ref{eq11}) for $\phi=\phi(r)$ becomes
 \be{eq21}
 \nabla^2\phi=\frac{d^2\phi}{dr^2}+\frac{2}{r}\frac{d\phi}{dr}
 =-V_{,\phi}-A_{,\phi}\,\rho=-V_{\rm eff,\phi}\,,
 \ee
 where $\rho(r)=\rho_c$ for $r<R_c$ and $\rho(r)=\rho_\infty$ for $r>R_c\,$.
 It is worth noting that at $r=R_c\,$, the matter density $\rho$ jumps
 from $\rho_c$ to $\rho_\infty$. Accordingly, the effective potential
 $V_{\rm eff}$ also undergoes a jump, while its shape changes. Noting
 $\phi_{\rm max}\propto\rho^{1/(s-1)}$ from Eq.~(\ref{eq14}) and $1<s<2$,
 $\phi_{\rm max}$ decreases when the matter density $\rho$ jumps from
 $\rho_c$ to $\rho_\infty$. This means that $V_{\rm eff}$ jumps to the
 left side, in contrast to the case of chameleon
 mechanism~\cite{Khoury:2003aq,Khoury:2003rn}. We present the
 demonstrational plots of $V_{\rm eff}$ in Fig.~\ref{fig1} (not to scale).
 We denote the field values minimizing $V_{\rm eff}$ as $\phi_c$ and
 $\phi_\infty$ for $r<R_c$ and $r>R_c\,$, respectively. Eq.~(\ref{eq21})
 is a second order differential equation, and hence two boundary conditions
 are needed. Following~\cite{Khoury:2003aq,Khoury:2003rn}, we require
 that the solution should be non-singular at the origin,
 \be{eq22}
 d\phi/dr=0\quad {\rm at}\quad r=0\,,
 \ee
 and the fifth-force tends to zero ($d\phi/dr\to 0$, n.b. Eq.~(\ref{eq5})
 still holds in the inverse chameleon mechanism) as $r\to\infty$. The
 latter is actually equivalent to
 \be{eq23}
 \phi\to\phi_\infty
 \quad {\rm as}\quad r\to\infty\,,
 \ee
 which is natural since $\rho=\rho_\infty$ at infinity.

Following e.g.~\cite{Khoury:2003aq,Khoury:2003rn}, to get an intuition, it
 is useful to think of $r$ as a ``\,time\,'' coordinate and $\phi$ as the
 ``\,position\,'' of a ``\,particle\,'', treating Eq.~(\ref{eq21}) as a
 dynamical problem in classical mechanics. In this language, $d^2\phi/dr^2$
 and $d\phi/dr$ are the ``\,acceleration\,'' and the ``\,speed\,'',
 respectively, while $-V_{\rm eff,\phi}$ is a ``\,time-dependent force\,''
 and $(2/r)\,d\phi/dr$ is a ``\,speed-dependent damping term\,'' (or
 ``\,friction\,''). At the initial ``\,time\,'' $r=0$, the ``\,particle\,''
 is at rest (see Eq.~(\ref{eq22})), and begins from the initial value
 $\phi_i\equiv\phi(r=0)$. For small $r$, the ``\,damping term\,'' ($\propto
 1/r$) is large, and hence the ``\,particle\,'' is frozen at $\phi=\phi_i$
 for a long ``\,time\,'' ($0<r<R_{roll}$). The frozen ``\,time\,'' (i.e. the
 value of $R_{roll}$) depends on the slope of the potential, namely the
 ``\,driving term\,'' $-V_{\rm eff,\phi}$. Note that $V_{\rm eff,\phi}\simeq
 0$ for a $\phi_i$ close to $\phi_c\,$, but it is large enough for a
 $\phi_i$ sufficiently displaced from $\phi_c\,$. In fact, they correspond
 to the cases of ``\,thin shell\,'' and ``\,thick shell\,'', respectively.
 As $r$ increases, the ``\,damping term\,'' decreases. Finally, at $r\simeq
 R_{roll}$, the ``\,damping term\,'' becomes smaller than the ``\,driving
 term\,'' $-V_{\rm eff,\phi}$, and then the ``\,particle\,'' begins to roll
 down the effective potential $V_{\rm eff}$, as shown in the right panel
 of Fig.~\ref{fig1}. Later, it arrives at $r=R_c\,$, where $V_{\rm eff}$
 suddenly changes as $\rho$ jumps from $\rho_c$ to $\rho_\infty$. But $\phi$
 and $d\phi/dr$ are continuous at $r=R_c\,$.

Outside the compact object, the ``\,particle\,'' changes to climb up the new
 $V_{\rm eff}$ as shown in the left panel of Fig.~\ref{fig1}. At this stage,
 the ``\,force\,'' $-V_{\rm eff,\phi}$ changes its sign, and help the
 ``\,friction\,'' to pull back the ``\,particle\,''. But the ``\,speed\,''
 of the ``\,particle\,'' is large enough compared to the new slope of the
 effective potential $V_{\rm eff,\phi}$, and hence it keeps moving leftwards
 by inertia. At the first stage outside the compact object ($R_c<r<R_t$),
 because the ``\,friction\,'' $(2/r)\,d\phi/dr$ is much larger than
 $-V_{\rm eff,\phi}$ due to the relatively large ``\,speed\,'' $d\phi/dr$,
 the ``\,force\,'' $-V_{\rm eff,\phi}$ can be completely neglected (although
 $-V_{\rm eff,\phi}\not\approx 0$ if $\phi(r=R_c)$ is sufficiently far from
 $\phi_\infty$, as shown in the left panel of Fig.~\ref{fig1}). We call
 $R_c<r<R_t$ the transition region. As $r$ increases, the ``\,friction\,''
 $(2/r)\,d\phi/dr$ becomes small, because the ``\,speed\,'' $d\phi/dr$ is
 decelerated and $2/r$ decreases. Meanwhile, the slope of $V_{\rm eff,\phi}$
 also tends to $0$, as the ``\,particle\,'' climbs up $V_{\rm eff}$ outside
 the compact object, as shown in the left panel of Fig.~\ref{fig1}. At the
 second stage ($r>R_t$), $\phi$ becomes fairly close to $\phi_\infty$, while
 $-V_{\rm eff,\phi}\simeq 0$ indeed. In the end, it will reach $\phi_\infty$
 as $r\to\infty$. At this stage, the ``\,acceleration\,'' $d^2\phi/dr^2$ and
 the ``\,speed\,'' $d\phi/dr$ tend to $0$ (accordingly the ``\,friction\,''
 $(2/r)\,d\phi/dr$ also tends to $0$), so that the first order approximation
 of $-V_{\rm eff,\phi}\simeq 0$, namely $-V_{\rm eff,\phi\phi}(\phi-
 \phi_\infty)=m_\infty^2(\phi-\phi_\infty)$, makes sense. The width of
 transition region depends on the distance between $\phi(r=R_c)$ and
 $\phi_\infty$. In the case of thin shell mentioned above, we will see that
 $|(\phi_c-\phi_\infty)/(6\beta M_{pl}\Phi_c)|\ll 1$ in the following (as in
 the case of chameleon mechanism~\cite{Khoury:2003aq,Khoury:2003rn}). So,
 $\phi(r=R_c)$ between $\phi_c$ and $\phi_\infty$ is close enough to
 $\phi_\infty$. Thus, the transition region can be ignored in the case of
 thin shell regime. On the contrary, there will be a considerable transition
 region in the case of thick shell regime.


\vspace{-10mm}  


\subsection{The thin shell regime}\label{sec4b}

As in the chameleon mechanism~\cite{Khoury:2003aq,Khoury:2003rn},
 we consider the ``\,thin shell\,'' regime and the ``\,thick shell\,''
 regime in our inverse chameleon mechanism, one by one. The thin shell
 regime is defined by $|\phi_i-\phi_c|\ll\phi_c\,$, namely $\phi_i$ is
 very close to $\phi_c\,$. As mentioned above, due to the large ``\,damping
 term\,'' ($\propto 1/r$) and hence the ``\,particle\,'' is frozen at
 $\phi=\phi_i$ for a long ``\,time\,''. So, in the frozen region we have
 \be{eq24}
 \phi(r)\simeq\phi_i\simeq\phi_c\quad {\rm for}\quad 0<r<R_{roll}\,.
 \ee
 When $r\sim R_{roll}$, $\phi$ is still near $\phi_c$ but begins to roll,
 because the ``\,damping term\,'' $(2/r)\,d\phi/dr$ becomes relatively
 smaller and the ``\,driving term\,'' $-V_{\rm eff,\phi}$ makes sense, as
 mentioned above. In the rolling region $R_{roll}<r<R_c\,$, the slope of
 $V(\phi)$ is much smaller than the one of $A(\phi)\,\rho$ as soon as
 $\phi$ is displaced significantly from $\phi_c\,$, similar to the case of
 chameleon mechanism~\cite{Khoury:2003aq,Khoury:2003rn} (one can see this
 by simply plotting $A(\phi)\,\rho$ and $V(\phi)$ in the same plane versus
 $\phi$). Thus, using $|V_{,\phi}|\ll A_{,\phi}\,\rho_c\,$, Eq.~(\ref{eq21})
 can be approximated by
 \be{eq25}
 \nabla^2\phi=\frac{d^2\phi}{dr^2}+\frac{2}{r}\frac{d\phi}{dr}
 \simeq -A_{,\phi}\,\rho_c\simeq -\frac{\beta\rho_c}{M_{pl}}
 \left(1+\frac{\beta\phi}{M_{pl}}+\cdots\right)\,,
 \ee
 where we have used $A(\phi)$ in Eq.~(\ref{eq13}), and considered its
 Taylor expansion up to the first order, noting $\beta\phi/M_{pl}\ll 1$,
 similar to e.g.~\cite{Khoury:2003aq,Khoury:2003rn}. The approximate
 solution of Eq.~(\ref{eq25}) is given by
 \be{eq26}
 \phi(r)\simeq -\frac{\beta\rho_c}{3M_{pl}}\left(\frac{r^2}{2}+
 \frac{\hat{c}}{r}\right)+\bar{c}+\phi_c\,.
 \ee
  Note that $\phi$ and $d\phi/dr$ are continuous at $r=R_{roll}$. Requiring
 $d\phi/dr=0$ and $\phi=\phi_c$ at $r=R_{roll}$, we find that
 $\hat{c}=R_{roll}^3$ and $\bar{c}=\beta\rho_c\,R_{roll}^2/(2M_{pl})$.
 So, in the rolling region we obtain
 \be{eq27}
 \phi(r)\simeq -\frac{\beta\rho_c}{3M_{pl}}\left(\frac{r^2}{2}+
 \frac{R_{roll}^3}{r}\right)+\frac{\beta\rho_c\,R_{roll}^2}{2M_{pl}}+
 \phi_c\quad {\rm for}\quad R_{roll}<r<R_c\,.
 \ee
 It is worth noting that the approximation of separating the solution for
 $0<r<R_c$ into Eqs.~(\ref{eq24}) and (\ref{eq27}) makes sense only if
 $R_c-R_{roll}\ll R_c$ (namely the shell is thin), for otherwise there is
 no clear separation between the two regions, and one needs a solution valid
 over the entire range $r<R_c$~\cite{Khoury:2003aq,Khoury:2003rn}. Then, the
 ``\,particle\,'' arrives at $r=R_c\,$, where the matter density $\rho$
 jumps from $\rho_c$ to $\rho_\infty$, and $\phi$ changes to climb up the
 new $V_{\rm eff}$ with $\rho=\rho_\infty$, as shown in Fig.~\ref{fig1}. As
 mentioned above, in the thin shell regime, the transition region can be
 ignored, because $\phi(r=R_c)$ is fairly close to $\phi_\infty$. So,
 $-V_{\rm eff,\phi}\simeq 0$, and its first order approximation makes sense.
 Noting that $\phi$ is very close to $\phi_\infty$, Eq.~(\ref{eq21}) becomes
 \be{eq28}
 \nabla^2\phi=\frac{d^2\phi}{dr^2}+\frac{2}{r}\frac{d\phi}{dr}=
 -V_{\rm eff,\phi}=-V_{\rm eff,\phi}(\phi_\infty)-V_{\rm eff,\phi\phi}
 (\phi_\infty)\left(\phi-\phi_\infty\right)+\dots\simeq
 m_\infty^2\left(\phi-\phi_\infty\right)\,,
 \ee
 where we have used Eq.~(\ref{eq12}). Its solution is given by
 \be{eq29}
 \phi(r)\simeq\frac{C_\infty\,e^{-m_\infty(r-R_c)}}{r}+\phi_\infty\,,
 \ee
 in which the divergent solution $\exp(m_\infty(r-R_c))/r$ has been excluded
 by the boundary condition in Eq.~(\ref{eq23}). The two unknowns $C_\infty$
 and $R_{roll}$ can be determined by requiring $d\phi/dr$ and $\phi$ are
 continuous at $r=R_c\,$. Matching $d\phi/dr$ from Eqs.~(\ref{eq27}) and
 (\ref{eq29}) at $r=R_c$ gives
 \be{eq30}
 \frac{\beta\rho_c}{3M_{pl}}\left(R_c-\frac{R_{roll}^3}{R_c^2}\right)
 =\frac{C_\infty}{R_c^2}\left(1+m_\infty R_c\right)\,.
 \ee
 In the case of chameleon mechanism~\cite{Khoury:2003aq,Khoury:2003rn},
 $m_\infty R_c\ll 1$ because the mass of the scalar field is small when
 the local matter density $\rho_\infty$ is low. However, in the inverse
 chameleon mechanism, $m_\infty R_c\gg 1$ because the mass of the scalar
 field is large when the local matter density $\rho_\infty$ is low. This
 is a key difference between these two mechanisms. We can look at this
 point carefully. In the case of thin shell, the fifth-force range $m_c^{-1}
 \ll R_c\,$, while $m_\infty>m_c$ for $\rho_\infty<\rho_c\,$, since
 $m_\phi$ is a decreasing function of $\rho$ in the inverse chameleon
 mechanism, as mentioned in Sec.~\ref{sec3a}. Thus, $m_\infty^{-1}<m_c^{-1}
 \ll R_c$ and hence $m_\infty R_c\gg 1$. So, in our case, from
 Eq.~(\ref{eq30}) we find that
 \be{eq31}
 C_\infty\simeq\frac{\beta\rho_c}{3M_{pl}}\frac{R_c^2}{m_\infty}\left(1-
 \frac{R_{roll}^3}{R_c^3}\right)=\frac{\beta}{4\pi M_{pl}}\frac{M_c}
 {m_\infty R_c}\left(1-\frac{R_{roll}^3}{R_c^3}\right)\simeq
 \frac{\beta}{4\pi M_{pl}}\frac{M_c}{m_\infty R_c}\left(\frac{3\Delta R_c}
 {R_c}\right)\,,
 \ee
 where we have used
 \be{eq32}
 \frac{\Delta R_c}{R_c}=\frac{R_c-R_{roll}}{R_c}\ll 1\,,
 \ee
 as mentioned above. On the other hand, matching $\phi$ in Eqs.~(\ref{eq27})
 and (\ref{eq29}) at $r=R_c$ gives
 \be{eq33}
 C_\infty\simeq\left(\phi_c-\phi_\infty\right)R_c\,.
 \ee
 The two $C_\infty$ in Eqs.~(\ref{eq31}) and (\ref{eq33}) must be equal.
 Introducing the Newtonian potential at the surface of the object $\Phi_c
 \equiv -GM_c/R_c<0$, we have
 \be{eq34}
 \frac{\Delta R_c}{R_c}=\frac{\phi_c-\phi_\infty}{6\beta
 M_{pl}\,|\Phi_c|}\cdot m_\infty R_c\ll 1\,,
 \ee
 which means that
 \be{eq35}
 \frac{\phi_c-\phi_\infty}{6\beta M_{pl}\,|\Phi_c|}\ll\frac{1}
 {m_\infty R_c}\ll 1\,,
 \ee
 since $m_\infty R_c\gg 1$ as mentioned above. Substituting
 Eq.~(\ref{eq31}) into Eq.~(\ref{eq29}), we find the exterior solution
 \be{eq36}
 \phi(r)\simeq\frac{\beta}{4\pi M_{pl}}\frac{1}{m_\infty R_c}
 \left(\frac{3\Delta R_c}{R_c}\right)\frac{M_c\,e^{-m_\infty(r-R_c)}}{r}
 +\phi_\infty\quad {\rm for}\quad r>R_c\,.
 \ee
 Clearly, there are double suppressions $\Delta R_c/R_c\ll 1$
 and $1/(m_\infty R_c)\ll 1$ before the Yukawa-suppression in
 Eq.~(\ref{eq36}). Thus, $\phi\simeq\phi_\infty$ soon after $r\gtrsim
 R_c\,$, and hence the fifth-force mediated by the scalar field $\phi$
 is nearly zero. We can see the thin-shell effect on the other hand. As
 mentioned above, Eq.~(\ref{eq5}) still holds in the inverse chameleon
 mechanism. The strength of fifth-force is characterized by
 \be{eq37}
 \boldsymbol{a}_\phi=-\nabla\ln A(\phi)=-\frac{\beta}{M_{pl}}\nabla\phi\,,
 \ee
 while the strength of gravitational force is characterized by
 $\boldsymbol{a}_g=-\nabla\Phi_N$. Noting that $\Phi_c$ is the Newtonian
 potential at the surface of the object, Eq.~(\ref{eq35}) indicates that
 the fifth-force is extremely smaller than the gravitational force. The
 physical reason for the thin-shell effect has been mentioned at the end
 of Sec.~\ref{sec2}. That is, the fifth-force range $m_c^{-1}\ll R_c\,$,
 and hence the contribution from the core to the exterior profile is
 significantly suppressed. Only the contribution from a thin shell beneath
 the surface contributes considerably to the exterior profile. On the
 other hand, gravity couples to the entire bulk of the object. Thus,
 the fifth-force mediated by the scalar field $\phi$ on an exterior
 test particle is suppressed compared to the gravitational force. The
 thin-shell effect is the key to evade the fifth-force tests.


\subsection{The thick shell regime}\label{sec4c}

Let us turn to the thick shell regime. In this case, $\phi_i\lesssim
 \phi_c\,$, namely the scalar field at $r=0$ is sufficiently displaced
 from $\phi_c\,$. There is no ``\,friction-dominated\,'' region, since the
 ``\,driving term\,'' $-V_{\rm eff,\phi}$ is large at $\phi_i$, as shown
 in the right panel of Fig.~\ref{fig1}. So, the ``\,particle\,'' begins to
 roll almost as soon as it is released at $r=0$. Similar to the case of
 chameleon mechanism~\cite{Khoury:2003aq,Khoury:2003rn}, the interior
 solution can be obtained by taking the $R_{roll}\to 0$ limit
 of Eq.~(\ref{eq27}) and replacing $\phi_c$ by $\phi_i$, namely
 \be{eq38}
 \phi(r)\simeq -\frac{\beta\rho_c\,r^2}{6M_{pl}}+\phi_i
 \quad {\rm for}\quad 0<r<R_c\,.
 \ee
 At $r=R_c\,$, the matter density $\rho$ jumps from $\rho_c$ to
 $\rho_\infty$, and $\phi$ changes to climb up the new $V_{\rm eff}$
 with $\rho=\rho_\infty$, as shown in the left panel of Fig.~\ref{fig1}.
 As mentioned above, there is a considerable transition region in the
 case of thick shell regime, because $\phi(r=R_c)$ can be sufficiently
 far from $\phi_\infty$, unlike the thin shell regime. As mentioned
 in the last paragraph of Sec.~\ref{sec4a}, in the transition region
 $R_c<r<R_t$, $-V_{\rm eff,\phi}\not\approx 0$, but it is much less than
 the ``\,friction\,'' $(2/r)\,d\phi/dr$ due to the relatively large
 ``\,speed\,'' $d\phi/dr$. In this case, Eq.~(\ref{eq21}) can
 be approximated by
 \be{eq39}
 \nabla^2\phi=\frac{d^2\phi}{dr^2}+\frac{2}{r}\frac{d\phi}{dr}\simeq 0\,,
 \ee
 while $-V_{\rm eff,\phi}\not\approx 0$ (including its first order term
 $\propto -V_{\rm eff,\phi\phi}$) is completely neglected. Its solution
 reads
 \be{eq40}
 \phi(r)\simeq\frac{C_t}{r}+\phi_t\,,
 \ee
 where $C_t$ and $\phi_t$ are both integration constants, which can be
 determined by requiring $d\phi/dr$ and $\phi$ are continuous at $r=R_c\,$.
 Matching $d\phi/dr$ from Eqs.~(\ref{eq38}) and (\ref{eq40}) at $r=R_c$
 gives
 \be{eq41}
 C_t=\frac{\beta\rho_c\, R_c^3}{3M_{pl}}=\frac{\beta M_c}{4\pi M_{pl}}\,.
 \ee
 So, the solution in the transition region is given by
 \be{eq42}
 \phi(r)\simeq\frac{\beta}{4\pi M_{pl}}\frac{M_c}{r}+\phi_t
 \quad {\rm for}\quad R_c<r<R_t\,.
 \ee
 Matching $\phi$ in Eqs.~(\ref{eq38}) and (\ref{eq42}) at $r=R_c$ leads to
 \be{eq43}
 \phi_i-\phi_t=3\beta M_{pl}\,|\Phi_c|\,,
 \ee
 and hence the integration constant $\phi_t$ is known. As mentioned
 in the last paragraph of Sec.~\ref{sec4a}, at $r>R_t$, $\phi$ becomes
 fairly close to $\phi_\infty$, while $-V_{\rm eff,\phi}\simeq 0$ indeed. At
 this stage, the ``\,acceleration\,'' $d^2\phi/dr^2$ and the ``\,speed\,''
 $d\phi/dr$ tend to $0$ (accordingly the ``\,friction\,'' $(2/r)\,d\phi/dr$
 also tends to $0$), so that the first order approximation of $-V_{\rm
 eff,\phi}\simeq 0$, i.e.~$-V_{\rm eff,\phi\phi}(\phi-\phi_\infty)=
 m_\infty^2(\phi-\phi_\infty)$, makes sense. Noting that $\phi$ is very
 close to $\phi_\infty$, Eq.~(\ref{eq21}) becomes
 \be{eq44}
 \nabla^2\phi=\frac{d^2\phi}{dr^2}+\frac{2}{r}\frac{d\phi}{dr}=
 -V_{\rm eff,\phi}=-V_{\rm eff,\phi}(\phi_\infty)-V_{\rm eff,\phi\phi}
 (\phi_\infty)\left(\phi-\phi_\infty\right)+\dots\simeq
 m_\infty^2\left(\phi-\phi_\infty\right)\,,
 \ee
 whose solution (satisfying the boundary condition in Eq.~(\ref{eq23})) is
 given by
 \be{eq45}
 \phi(r)\simeq\frac{C_\infty\,e^{-m_\infty(r-R_c)}}{r}+\phi_\infty\,.
 \ee
 The two unknowns $C_\infty$ and $R_t$ can be determined by
 requiring $d\phi/dr$ and $\phi$ are continuous at $r=R_t$. Matching
 $\phi$ in Eqs.~(\ref{eq42}) and (\ref{eq45}) at $r=R_t$ gives
 \be{eq46}
 \frac{\beta}{4\pi M_{pl}}\frac{M_c}{R_t}+\phi_t\simeq\phi_\infty\,,
 \ee
 where we have neglected the first term in the right hand side
 of Eq.~(\ref{eq45}), because $\phi(r=R_t)$ is very close to $\phi_\infty$
 by definition of $R_t$. From Eq.~(\ref{eq46}), it is easy to see that
 $\phi_t<\phi_\infty$ is on the left side of $\phi_\infty$. This has no
 problem since $\phi_t$ is just an integration constant without special
 meaning in physics, as mentioned above. Substituting Eq.~(\ref{eq43})
 into Eq.~(\ref{eq46}), we have
 \be{eq47}
 R_t=\frac{2}{3}\,R_c
 \left(1-\frac{\phi_i-\phi_\infty}{3\beta M_{pl}\,|\Phi_c|}\right)^{-1}\,.
 \ee
 Noting $R_t>R_c>0$, it is easy to see that
 \be{eq48}
 \frac{1}{3}<\frac{\phi_i-\phi_\infty}{3\beta M_{pl}\,|\Phi_c|}<1\,.
 \ee
 Clearly, the width of transition region $\Delta_t=R_t-R_c$ can
 be fairly large for suitable $\phi_i$. On the other hand,
 from Eq.~(\ref{eq48}) we find
 \be{eq49}
 \frac{\phi_c-\phi_\infty}{6\beta M_{pl}\,
 |\Phi_c|}>\frac{\phi_i-\phi_\infty}{6\beta M_{pl}\,|\Phi_c|}>\frac{1}{6}\,,
 \ee
 which implies that the thin shell condition in Eqs.~(\ref{eq34}) or
 (\ref{eq35}) does not hold in the thick shell regime, as expected. Matching
 $d\phi/dr$ from Eqs.~(\ref{eq42}) and (\ref{eq45}) at $r=R_t$ gives
 \be{eq50}
 C_\infty=\frac{\beta}{4\pi M_{pl}}\frac{M_c}
 {1+m_\infty R_t}\,e^{-m_\infty(R_c-R_t)}\,.
 \ee
 Substituting Eq.~(\ref{eq50}) into Eq.~(\ref{eq45}), we have
 the exterior solution
 \be{eq51}
 \phi(r)\simeq\frac{\beta}{4\pi M_{pl}}\frac{M_c}
 {1+m_\infty R_t}\frac{e^{-m_\infty(r-R_t)}}{r}+\phi_\infty
 \quad {\rm for}\quad r>R_t\,.
 \ee
 Note that $m_\infty$ is fairly large if the background density
 $\rho_\infty$ is fairly low, and hence there is another suppression
 factor $1/(1+m_\infty R_t)\simeq 1/(m_\infty R_t)\ll 1$ before the
 Yukawa-suppression in Eq.~(\ref{eq51}). Thus, $\phi\simeq\phi_\infty$ soon
 after $r\gtrsim R_t$, and hence the fifth-force mediated by the scalar
 field $\phi$ is nearly zero (n.b. Eq.~(\ref{eq37})). This can be easily
 understood in physics. In the case of thick shell regime, the fifth-force
 range $m_c^{-1}\gg R_c\,$, and hence the entire bulk of the object
 contributes significantly to the exterior profile. Although the fifth-force
 range $m_c^{-1}\gg R_c\,$, it is still finite. So, the contribution from
 the object extends up to $r\sim R_t$ at the most, but it is significantly
 suppressed at $r\gtrsim R_t$. The interesting region is $r<R_t$. In the
 transition region $R_c<r<R_t$, the corresponding $\phi(r)$ solution is
 given by Eq.~(\ref{eq42}), while the Newtonian potential $\Phi_{\rm N}
 \simeq -GM_c/r$ from Eq.~(\ref{eq6}) with $A(\phi)\simeq 1$ for $\beta
 \phi/M_{pl}\ll 1$. Substituting them into Eq.~(\ref{eq5}), we obtain the
 acceleration felt by a test particle in the transition region $R_c<r<R_t$,
 \be{eq52}
 \boldsymbol{a}=-\nabla\left(\Phi_{\rm N}+\ln A(\phi)\right)
 \simeq -\frac{GM_c}{r^2}\,\boldsymbol{e}_r-
 \frac{\beta}{M_{pl}}\frac{d\phi}{dr}\,\boldsymbol{e}_r
 \simeq -\frac{GM_c}{r^2}\,(1-2\beta^2)\,\boldsymbol{e}_r\,,
 \ee
 which means that the gravitational force is considerably weakened by the
 fifth-force mediated by the scalar field $\phi\,$. It is equivalent to a
 weakened gravitational force with $G_{\rm eff}=(1-2\beta^2)\,G$. So, it
 is possible to test the fifth-force in the transition region outside the
 object in thick shell regime, such as compact stars including white dwarfs
 and neutron stars. On the other hand, inside the object $r<R_c\,$, the
 interior $\phi(r)$ solution is given by Eq.~(\ref{eq38}), while the
 Newtonian potential $\Phi_{\rm N}\simeq -\left(2\pi G\rho_c/3\right)(3R_c^2
 -r^2)$ from Eq.~(\ref{eq6}) with $A(\phi)\simeq 1$ for $\beta\phi/M_{pl}
 \ll 1$. Substituting them into Eq.~(\ref{eq5}), we obtain the acceleration
 felt by a test particle inside the object $r<R_c\,$, namely
 \be{eq53}
 \boldsymbol{a}=-\nabla\left(\Phi_{\rm N}+\ln A(\phi)\right)
 \simeq -\frac{4\pi G\rho_c}{3}\,\boldsymbol{r}-
 \frac{\beta}{M_{pl}}\frac{d\phi}{dr}\,\boldsymbol{e}_r\simeq
 -\frac{4\pi G\rho_c}{3}\,(1-2\beta^2)\,\boldsymbol{r}\,.
 \ee
 It is also equivalent to a weakened gravitational force with
 $G_{\rm eff}=(1-2\beta^2)\,G$ inside the object in thick shell regime. So,
 it is possible to increase the mass limits for compact stars such as white
 dwarfs and neutron stars. Noting that $\beta$ can be ${\cal O}(1)$,
 the effect of fifth-force could be fairly significant.


\section{Mass limits for compact stars in the inverse chameleon
 mechanism}\label{sec5}


\subsection{White dwarfs}\label{sec5a}

Here, we consider the mass limits for compact stars in the inverse chameleon
 mechanism. As shown above, the gravitational force is considerably weakened
 by the fifth-force mediated by the scalar field $\phi$ inside a compact
 object in thick shell regime. However, the object is assumed to be
 homogeneous in Sec.~\ref{sec4}, since its main goal is to show how to evade
 the fifth-force tests on earth and in solar system where a homogeneous
 object is a good enough approximation. As is well known, compact stars such
 as white dwarfs and neutron stars are highly inhomogeneous. So, we cannot
 directly use the results of Sec.~\ref{sec4}, for example, $G_{\rm eff}=
 (1-2\beta^2)\,G$. Instead, here we should consider compact stars in
 general, without assuming homogeneousness.

At first, we consider white dwarfs. As is well known, one can use the
 Newtonian approximation for the calculation of white dwarf structure
 \cite{Camenzind:2007,Shapiro:2004,Glendenning:1996}. For a spherical star,
 the mass interior to a radius $r$ is given by
 \be{eq54}
 M(r)=\int_0^r 4\pi\tilde{r}^2\rho(\tilde{r})\,d\tilde{r}\,,
 \quad {\rm or}\quad \frac{dM(r)}{dr}=4\pi r^2\rho(r)\,.
 \ee
 We assume that the star is in a hydrostatic equilibrium. We consider an
 infinitesimal fluid element lying between $r$ and $r+dr$, which has an
 area $dS$ perpendicular to the radial direction, and a mass $dm$. The net
 outward pressure force on $dm$ is
 \be{eq55}
 P(r)\,dS-P(r+dr)\,dS=-dP\,dS=-\frac{dP}{dr}\,dr\,dS\,.
 \ee
 On the other hand, using Eq.~(\ref{eq5}), the gravitational force and the
 fifth-force mediated by the scalar field $\phi$ on $dm$ is given by
 \be{eq56}
 \boldsymbol{a}\,dm=-\nabla\left(\Phi_{\rm N}+\ln A\right)dm\,,
 \ee
 while $-\nabla\Phi_{\rm N}\simeq -(GM(r)/r^2)\,\boldsymbol{e}_r$ from
 Eq.~(\ref{eq6}) with $A(\phi)\simeq 1$ for $\beta\phi/M_{pl}\ll 1$, and
 $M(r)$ is given by Eq.~(\ref{eq54}), $A(\phi)$ is given by
 Eq.~(\ref{eq13}). Thus, in equilibrium we have
 \be{eq57}
 -\frac{dP}{dr}\,dr\,dS=\left(\frac{GM(r)}{r^2}+
 \frac{\beta}{M_{pl}}\frac{d\phi}{dr}\right)dm\,,
 \ee
 which is equivalent to
 \be{eq58}
 \frac{dP}{dr}=-\rho\left(\frac{GM(r)}{r^2}+
 \frac{\beta}{M_{pl}}\frac{d\phi}{dr}\right)\,.
 \ee
 Using Eq.~(\ref{eq54}), we can recast Eq.~(\ref{eq58}) as
 \be{eq59}
 \frac{1}{r^2}\frac{d}{dr}\left(\frac{r^2}{\rho}\frac{dP}{dr}\right)=
 -4\pi G\rho-\frac{\beta}{M_{pl}}\nabla^2\phi\,.
 \ee
 Substituting Eq.~(\ref{eq11}) or Eq.~(\ref{eq21}) into Eq.~(\ref{eq59}),
 and noting $|V_{,\phi}|\ll A_{,\phi}\,\rho$ inside the object in thick
 shell regime (as mentioned in Sec.~\ref{sec4}), we obtain (see
 also the note in~\cite{LEeq} for an alternative derivation)
 \be{eq60}
 \frac{1}{r^2}\frac{d}{dr}\left(\frac{r^2}{\rho}\frac{dP}{dr}\right)
 =-4\pi G\rho\,(1-2\beta^2)=-4\pi G_{\rm eff}\,\rho\,,
 \ee
 where we have used $\beta\phi/M_{pl}\ll 1$. Now, we arrive at the same
 position of the usual calculation of white dwarf structure
 \cite{Camenzind:2007,Shapiro:2004,Glendenning:1996} but with an effective
 gravitational constant
 \be{eq61}
 G_{\rm eff}=(1-2\beta^2)\,G\,.
 \ee
 Notice that the above derivations hold for the general $\rho=\rho(r)$,
 without assuming homogeneousness. Following e.g.~\cite{Camenzind:2007,
 Shapiro:2004,Glendenning:1996,Chandrasekhar:1939}, we can easily derive
 the mass limit for white dwarfs. We consider a polytropic equation
 of state (EOS) for the fermion gas (the electron gas),
 \be{eq62}
 P=K\rho^{\,\Gamma}=K\rho^{1+1/n}\,,
 \ee
 where $K$, $n$, $\Gamma=1+1/n$ are constants, and $n$ is the so-called
 polytropic index. It is convenient to introduce the dimensionless variables
 $\theta$ and $\xi$ by the parameterizations
 \be{eq63}
 \rho=\rho_0\,\theta^n\,,\quad r=\alpha\,\xi\,,
 \ee
 where $\rho_0\equiv\rho(r=0)$ is the central density, and
 \be{eq64}
 \alpha^2
 \equiv\frac{(n+1)\,K\rho_0^{1/n-1}}{4\pi G_{\rm eff}}\,.
 \ee
 Using these dimensionless variables, the hydrostatic equilibrium equation
 (\ref{eq60}) can be recast as the well-known Lan\'e-Emden equation
 \cite{Camenzind:2007,Shapiro:2004,Glendenning:1996,Chandrasekhar:1939}
 \be{eq65}
 \frac{1}{\xi^2}\frac{d}{d\xi}\left(\xi^2\frac{d\theta}{d\xi}
 \right)=-\theta^n\,.
 \ee
 It can be numerically solved with the boundary conditions
 at the center, namely
 \be{eq66}
 \theta=1\,,\quad \theta^\prime=0\quad {\rm at}\quad \xi=0\,,
 \ee
 where $\theta^\prime=d\theta/d\xi$. The surface of the star (where
 $P=\rho=0$) is located at $\theta(\xi_\ast)=0$. Eq.~(\ref{eq65}) can be
 integrated numerically, starting at $\xi=0$ with the boundary conditions
 in Eq.~(\ref{eq66}). For $n<5$, the solutions decrease monotonically and
 have a zero at a finite value $\xi_\ast$. For various polytropic EOS, the
 corresponding $\xi_\ast$ and $\xi_\ast^2\,|\theta^\prime(\xi_\ast)|$
 can be found in this way. For example, in the non-relativistic case
 ($\Gamma=5/3$ or $n=3/2$), one find~\cite{Camenzind:2007,Shapiro:2004,
 Glendenning:1996,Chandrasekhar:1939}
 \be{eq67}
 \xi_\ast=3.6537\,,\quad \xi_\ast^2\,|\theta^\prime(\xi_\ast)|=2.71406\,.
 \ee
 In the extreme relativistic case ($\Gamma=4/3$ or $n=3$), they
 are~\cite{Camenzind:2007,Shapiro:2004,Glendenning:1996,Chandrasekhar:1939}
 \be{eq68}
 \xi_\ast=6.89685\,,\quad \xi_\ast^2\,|\theta^\prime(\xi_\ast)|=2.01824\,.
 \ee
 With $\xi_\ast$, we obtain the stellar radius $R_\ast$ as a function
 of the central density
 \be{eq69}
 R_\ast=\alpha\,\xi_\ast=\sqrt{\frac{(n+1)\,K}{4\pi G_{\rm eff}}}\;
 \rho_0^{(1-n)/(2n)}\,\xi_\ast\,,
 \ee
 and the stellar mass $M_\ast$ as a function of the central density
 \bea
 M_\ast &=&\int_0^{R_\ast} 4\pi r^2\rho\,dr=4\pi\alpha^3\rho_0
 \int_0^{\xi_\ast}\xi^2\theta^n d\xi=-4\pi\alpha^3\rho_0\int_0^{\xi_\ast}
 \frac{d}{d\xi}\left(\xi^2\frac{d\theta}{d\xi}\right)d\xi\nonumber \\[1mm]
 &=& 4\pi\alpha^3\rho_0\,\xi_\ast^2\,|\theta^\prime(\xi_\ast)|
 =4\pi\left[\frac{(n+1)\,K}{4\pi G_{\rm eff}}\right]^{3/2}
 \rho_0^{(3-n)/(2n)}\,\xi_\ast^2\,|\theta^\prime(\xi_\ast)|\,.\label{eq70}
 \eea
 Eliminating the central density $\rho_0$ in Eqs.~(\ref{eq69})
 and (\ref{eq70}), we obtain the mass-radius relation as
 \be{eq71}
 M_\ast(R_\ast)=4\pi R_\ast^{(3-n)/(1-n)}\left[\frac{(n+1)\,K}
 {4\pi G_{\rm eff}}\right]^{n/(n-1)}\xi_\ast^2\,|\theta^\prime(\xi_\ast)|
 \,\xi_\ast^{(3-n)/(1-n)}\,.
 \ee
 For various polytropic EOS, the corresponding $K$ have been given in
 e.g.~\cite{Camenzind:2007,Shapiro:2004,Glendenning:1996,
 Chandrasekhar:1939}. We are interested in the extreme relativistic case
 ($\Gamma=4/3$ or $n=3$) which gives the mass limit for white dwarfs. Noting
 Eq.~(\ref{eq71}), the stellar mass $M_\ast$ is independent of radius
 $R_\ast$ in this case, namely
 \be{eq72}
 M_\ast=\left(1-2\beta^2\right)^{-3/2}M_{\rm Ch}=\left(1-
 2\beta^2\right)^{-3/2}\,1.457\,M_\odot\left(\frac{2}{\mu_e}\right)^2\,,
 \ee
 where we have used Eq.~(\ref{eq61}), and $\mu_e$ is the mean molecular
 weight per electron (usually $\mu_e\simeq 2$ for white dwarfs, but it is
 larger for different chemical compositions). From Eq.~(\ref{eq72}), it
 is easy to see that the mass limit for white dwarfs becomes
 $(1-2\beta^2)^{-3/2}$ times the well-known Chandrasekhar limit
 $M_{\rm Ch}$. On the other hand, from Eq.~(\ref{eq69}), the corresponding
 stellar radius $R_\ast$ is given by
 \be{eq73}
 R_\ast=\left(1-2\beta^2\right)^{-1/2}R_{\rm Ch}=\left(1-
 2\beta^2\right)^{-1/2}\,3.347\times 10^4\,{\rm km}\left(\frac{\rho_0}
 {10^6\,{\rm g/cm^3}}\right)^{-1/3}\left(\frac{2}{\mu_e}\right)^{2/3}\,,
 \ee
 which is also increased by a factor $(1-2\beta^2)^{-1/2}$. Of course, we
 should require $\beta^2<1/2$. Noting that $\beta$ can be ${\cal O}(1)$ in
 the inverse chameleon mechanism, the mass limit for white dwarfs could be
 significantly increased. For example, the mass limit for white dwarfs
 becomes about 1.66, 1.84, 2.15, 2.83, 3.95, 5.2, 6.75, 11.2 times
 the Chandrasekhar limit mass $M_{\rm Ch}$ for $\beta^2=1/7$, $1/6$, $1/5$,
 $1/4$, $0.3$, $1/3$, $0.36$, $2/5$, respectively. In fact, it can be larger
 than $3M_\odot$ for $\beta^2\gtrsim 1/5$, and hence the super-Chandrasekhar
 mass white dwarfs can be easily accommodated. In principle, the mass limit
 for white dwarfs can be very high for $\beta^2$ close enough to $1/2$.
 However, the value of $\beta^2$ will be constrained by observations (see
 discussions in Sec.~\ref{sec6}), and hence the mass limit for white dwarfs
 cannot be arbitrarily large in practice.


\subsection{Other compact stars}\label{sec5b}

In the case of white dwarfs, the Newtonian approximation is good enough.
 However, in the cases of neutron stars and other relativistic stars,
 the full relativistic hydrostatic equilibrium should be considered instead
 (see e.g.~\cite{Camenzind:2007,Shapiro:2004,Glendenning:1996}). The
 Tolman-Oppenheimer-Volkoff (TOV) equation is the corresponding master
 equation. On the other hand, the realistic (non-analytic) EOS should be
 considered, but which is not so clear to date. In the cases of neutron
 stars (and other relativistic stars), numerical computer codes are commonly
 employed. Therefore, it is not straightforward to obtain a simple factor
 increasing the mass limits (like the factor $(1-2\beta^2)^{-3/2}$ in the
 case of white dwarfs).

However, we argue that the mass limits for neutron stars and
 other relativistic stars will also be considerably increased in the inverse
 chameleon mechanism. At first, we have clearly shown that the fifth-force
 mediated by the scalar field $\phi$ will notably weaken the gravitational
 force inside the object in thick shell regime. The compact stars are
 stable due to the balance between gravitational force and degeneracy
 pressure. When the gravitational force is significantly weakened by the
 fifth-force, the same degeneracy pressure can of course support a much
 heavier mass. On the other hand, we note that compact stars are in a
 sequence, as is well known. If electron degeneracy pressure can support
 a white dwarf with a mass $M>3M_\odot\,$, $5M_\odot$ or even higher (for
 larger $\beta^2$), it certainly will not collapse into a neutron star or
 other relativistic stars such as quark star and gravastar. So, the mass
 limits for neutron stars and other relativistic stars must exceed the one
 for white dwarfs. While the mass limit for white dwarfs is significantly
 increased by a factor $(1-2\beta^2)^{-3/2}$, the mass limits for neutron
 stars and other relativistic stars can only be increased accordingly or
 even dramatically.


\section{Concluding remarks}\label{sec6}

As is well known, there are various mass limits for compact stars. For
 example, the maximum mass for non-rotating white dwarfs is given by the
 famous Chandrasekhar limit about $1.4 M_\odot\,$. Although the mass limit
 for neutron stars is not so clear to date, one of the widely accepted
 values is about $2.1 M_\odot\,$. Recently, challenges to these mass
 limits appeared. Motivated by the super-Chandrasekhar mass white dwarfs
 with masses up to $2.4 \sim 2.8 M_\odot\,$, and compact objects (probably
 neutron stars) in the mass gap (from $2.5 M_\odot$ or $3 M_\odot$ to
 $5 M_\odot$) inferred from gravitational waves detected by LIGO/Virgo in
 the third observing run (O3), we reconsider the mass limits for compact
 stars in the present work. Without invoking strong magnetic field and/or
 exotic EOS, we try to increase the mass limits for compact stars in
 modified gravity theory. In this work, we propose an inverse chameleon
 mechanism, and show that the fifth-force mediated by the scalar field can
 evade the severe tests on earth, in solar system and universe, but manifest
 itself in compact stars such as white dwarfs and neutron stars. The mass
 limits for compact stars in the inverse chameleon mechanism can be easily
 increased to $3 M_\odot\,$, $5 M_\odot$ or even larger.

In the literature, strong magnetic field and/or exotic EOS are frequently
 invoked in many scenarios to increase the mass limits for compact stars.
 On the other hand, the mass limits can also be increased for rigidly
 spinning compact stars (see e.g.~\cite{Halder:2020ahg}). However, the
 $2.6 M_\odot$ compact object found by LIGO/Virgo in GW190814 event has
 low primary spin~\cite{Abbott:2020khf}. As shown in this work,
 we consider that the scenarios employing modified gravity theories
 deserve further investigation.

In this work, a non-canonical scalar field (akin to phantom) is used in
 our inverse chameleon mechanism, and hence the gravitational force is
 weakened by the fifth-force mediated by the scalar field, n.b.~the
 effective gravitational constant $G_{\rm eff}=(1-2\beta^2)\,G$. In fact,
 one can instead consider a different inverse chameleon mechanism still
 using a canonical scalar field (akin to quintessence), and the action
 takes the same form given in Eq.~(\ref{eq1}), but $V(\phi)={\cal M}^{4-s}
 \,\phi^s$ with $1<s<2$, and $A(\phi)=\exp(-\beta\phi/M_{pl})$ with
 $\beta>0$. In this case, $m_\phi^2=V_{\rm eff,\phi\phi}(\phi_{\rm min})
 \propto\rho^{(s-2)/(s-1)}$ is also a decreasing function of the local
 matter density~$\rho$. Unfortunately, the gravitational force is instead
 strengthened by the fifth-force mediated by the scalar field, with
 $G_{\rm eff}=(1+2\beta^2)\,G$. So, it fails to increase the mass limits
 for compact stars. But we still mention it here with the hope to revive
 it for another completely different goal in the future.

Actually, one can see that $G_{\rm eff}=(1+2\beta^2)\,G$ in both cases
 of the original chameleon mechanism and the inverse chameleon mechanism
 with a canonical scalar field (akin to quintessence) mentioned above.
 In both cases, the fifth-force mediated by a canonical scalar field
 (akin to quintessence) is attractive, as is well known in quantum field
 theory (QFT). However, it is not the case of our inverse chameleon
 mechanism with a non-canonical scalar field (akin to phantom) proposed
 in the present work. In fact, it was shown in~\cite{Amendola:2004qb}
 that the phantom scalar field mediates a long-range repulsive force
 surprisingly. This is mainly due to the negative kinetic term of the
 phantom scalar field. So, it is easy to understand $G_{\rm eff}=(1-
 2\beta^2)\,G$ in our inverse chameleon mechanism, while $-2\beta^2$
 indicates the long-range repulsive fifth-force mediated by the
 non-canonical scalar field (akin to phantom).

The phantom scalar field with a negative kinetic term has led
 many interesting features (significantly different from the ones of
 canonical scalar field) to cosmology in the past two decades. On the
 other hand, it is worth noting that phantom dark energy whose
 EOS parameter $w<-1$ is slightly favored by the cosmological
 observations (e.g.~$w=-1.03\pm 0.03$ from the Planck 2018
 results~\cite{Aghanim:2018eyx}). Therefore, it is well motivated to also
 consider a non-canonical scalar field (akin to phantom) in astrophysics.
 Note that it was argued in e.g.~\cite{Cline:2003gs,Libanov:2007mq,
 Kaplinghat:2006jk} that phantom could avoid the quantum instability in the
 ultraviolet region. One can try to make the instability time scale greater
 than the age of the universe. However, phantom within a Lorentz invariant
 framework might be experimentally excluded~\cite{Cline:2003gs}. As is
 argued in e.g.~\cite{Cline:2003gs}, in order to keep the instability at
 unobservable levels, a Lorentz-violating ultraviolet cutoff $\Lambda$ must
 be applied to low-energy effective theories of phantom. It was found
 in~\cite{Cline:2003gs} that the cutoff $\Lambda$ is constrained by
 observations of the diffuse gamma-ray background, namely
 $\Lambda\lesssim 3\,{\rm MeV}$. As an explicit and simple example, we can
 consider a Lorentz-violating Lagrangian $\frac{1}{2}(\partial\phi)^2-
 \frac{1}{2}\Lambda^{-2}(\nabla^2 \phi)^2$ mentioned in~\cite{Cline:2003gs}
 (see also e.g.~\cite{Arkani-Hamed:2003pdi}). The second term is the key
 to keep the instability at unobservable levels, and it makes sense in
 the high-energy region above the cutoff $\Lambda$. On the other hand, in
 the low-energy region well below the cutoff $\Lambda$, the Lagrangian
 $\frac{1}{2}(\partial\phi)^2-\frac{1}{2}\Lambda^{-2}(\nabla^2 \phi)^2$
 effectively reduces to $\frac{1}{2}(\partial\phi)^2$, namely the one used
 in Eq.~(\ref{eq10}) of the present work. As is shown in Sec.~\ref{sec3b},
 the mass of phantom field $\phi$ is about ${\cal O}(10^{-3}\,{\rm eV})$ in
 the inverse chameleon mechanism, which is much lower than the cutoff
 $\Lambda\lesssim 3\,{\rm MeV}$. Thus, as a low-energy effective theory,
 one needs not worry about the quantum instability in the inverse chameleon
 mechanism. Note that Lorentz violation has been induced in many theories.
 For example, most theories of quantum gravity (QG) commonly predict that
 Lorentz violation might happen on high-energy scales. In fact, the
 observational hints for Lorentz violation were found in
 e.g.~\cite{Zou:2017ksd} by using the time-lag data of gamma-ray bursts
 (GRBs). Of course, the debate about quantum instability of phantom is still
 not completely settled in the literature by now, and we consider that it is
 better to keep an open mind to such kind of theories using a non-canonical
 scalar field (akin to phantom).

Note that $\beta$ is a constant in this work. So, the passage from $1-
 2\beta^2>0$ to $1-2\beta^2<0$ and vice versa cannot happen. Of course,
 it is interesting to consider a varying $\beta$ in some modified versions
 of the inverse chameleon mechanism, but this is beyond the scope of the
 present work. In principle, $1-2\beta^2<0$ can be allowed, and
 $G_{\rm eff}=(1-2\beta^2)\,G<0$ means that the repulsive fifth-force
 mediated by a non-canonical scalar field (akin to phantom) overcomes
 gravity. However, it is worth noting that $G_{\rm eff}=(1-2\beta^2)\,G$
 holds only in the cases of compact stars where the matter density is
 very high. As shown in this work, the inverse chameleon mechanism hides
 itself on earth, in solar system and universe, where the matter density
 is low. Thus, the usual stars (e.g. sun, stellar objects, planets, moons)
 and most objects in the universe can still be formed and exist as in GR.
 Nothing changes in the cases of low matter density even if $1-2\beta^2<0$.
 Only in the cases of very high matter density, the inverse chameleon
 mechanism manifests itself, and then $1-2\beta^2<0$ will prevent the
 existence of compact stars such as white dwarfs and neutron stars. On the
 contrary, the observational fact that white dwarfs and neutron stars do
 exist must require $1-2\beta^2>0$. In other words, $\beta^2<1/2$ must be
 constrained by the observations.

A natural question is how to test the inverse chameleon mechanism. On the
 other hand, in principle, the mass limits can be arbitrarily large by
 using $\beta^2\to 1/2$ in the factor $(1-2\beta^2)^{-3/2}$. Of course,
 this cannot happen in a reasonable theory. A constraint on $\beta^2$
 must be set from observations and/or experiments. Since the inverse
 chameleon mechanism hides itself on earth, in solar system and universe,
 it cannot be tested here. The inverse chameleon mechanism manifests itself
 in or near compact stars such as white dwarfs and neutron stars. The
 discovery of exoplanets shared the 2019 Nobel Prize in Physics. In fact,
 the first two exoplanets announced in 1992 are orbiting a pulsar (neutron
 star)~\cite{exoplanet}. To date, some exoplanets orbiting white dwarfs and
 neutron stars have been found~\cite{WDexoplanet,NSexoplanet,NASAexoplanet,
 Veras:2021}. As shown in Sec.~\ref{sec4c}, there is a transition region
 $R_c<r<R_t$ outside compact stars, where $G_{\rm eff}=(1-2\beta^2)\,G$.
 The exoplanets inside the transition region $R_c<r<R_t$ feels a weakened
 gravitational force, and hence their orbits will be affected. So, the
 observations of exoplanets orbiting compact stars such as white dwarfs and
 neutron stars might be used to test the inverse chameleon mechanism, and
 set a constraint on $\beta^2$. However, since the semi-major axes of the
 detected exoplanets orbiting white dwarfs and neutron stars are usually too
 large while the transition regions outside compact stars are not so wide,
 no considerable constraints can be made to date. We hope that an exoplanet
 very close to compact star can be found in the future. Another type of
 tests might come from gravitational waves. Two compact stars are very close
 in the last stage of their coalescence, so that they enter the transition
 regions of each other, where the gravitational force is notably weakened by
 the fifth-force mediated by the scalar field. In the last stage of the
 coalescence of binary neutron star, neutron star -- black hole, neutron
 star -- white dwarf, white dwarf -- black hole, and binary white dwarf, the
 inverse chameleon mechanism will affect these two compact stars at a very
 short distance. Thus, gravitational waves from the last stage of binary
 compact star coalescence might carry the information about the inverse
 chameleon mechanism. We encourage the GW community to search it in the
 GW data.


\section*{ACKNOWLEDGEMENTS}

We thank the anonymous referee for useful comments and suggestions,
 which helped us to improve this work. We are grateful to Zong-Kuan~Guo,
 Shupeng~Song, Shou-Long~Li, Jing-Yi~Jia, Da-Chun~Qiang, Hua-Kai~Deng
 and Han-Yue~Guo for kind help and discussions. This work was supported
 in part by NSFC under Grants No.~11975046 and No.~11575022.

\renewcommand{\baselinestretch}{1.1}



\begin{thebibliography}{99}
\bibitem{LEeq}
Alternatively, we can derive Eq.~(\ref{eq60}) by using the Poisson equation
 for the Newtonian potential $\Phi_{\rm N}$ given in Eq.~(\ref{eq6}). As
 is well known, $\nabla^2 f=(1/r^2)\,d\,(r^2\,df/dr)/dr$ in spherical
 coordinates for any $f(r)$. In equilibrium, the total force, namely
 Eq.~(\ref{eq55}) plus Eq.~(\ref{eq56}), is zero. Thus, we have $-(1/\rho)\,
 dP/dr=\nabla\,(\Phi_{\rm N}+\ln A)=d\Phi_{\rm N}/dr+(\beta/M_{pl})
 \,d\phi/dr$ for $A=\exp\left(\beta\phi/M_{pl}\right)$. Substituting
 $d\Phi_{\rm N}/dr=-(1/\rho)\,dP/dr-(\beta/M_{pl})\,d\phi/dr$
 into Eq.~(\ref{eq6}), and then using $\nabla^2\phi$ from Eq.~(\ref{eq11})
 or Eq.~(\ref{eq21}) with $|V_{,\phi}|\ll A_{,\phi}\,\rho\,$, we finally
 obtain Eq.~(\ref{eq60}) for $\beta\phi/M_{pl}\ll 1$. In this alternative
 way, no explicit solutions for $\Phi_{\rm N}$ and $\phi$ are required, and
 of course we need not assume that the object is homogeneous.
\end{thebibliography}
\end{document}